\documentclass{elsarticle}

\usepackage{lscape}

\newcommand{\captionfonts}{\normalsize}
\makeatletter  
\long\def\@makecaption#1#2{%
  \vskip\abovecaptionskip
  \sbox\@tempboxa{{\captionfonts #1: #2}}%
  \ifdim \wd\@tempboxa >\hsize
    {\captionfonts #1: #2\par}
  \else
    \hbox to\hsize{\hfil\box\@tempboxa\hfil}%
  \fi
  \vskip\belowcaptionskip}
\makeatother   

\begin{document}

\title{Observation of Markarian 421 in TeV gamma rays over a 14-year time span}

\author[a1,a2]{V.~A.~Acciari}

\author[a3]{T.~Arlen}

\author[a4]{T.~Aune}

\author[a2]{W.~Benbow}

\author[a5]{R.~Bird}

\author[a4]{A.~Bouvier}

\author[a6]{S.~M.~Bradbury}

\author[a7]{J.~H.~Buckley}

\author[a7]{V.~Bugaev}

\author[a8]{I.~de~la~Calle~Perez}

\author[a9]{D.~A.~Carter-Lewis}

\author[a10]{A.~Cesarini}

\author[a11]{L.~Ciupik}

\author[a5]{E.~Collins-Hughes}

\author[a10]{M.~P.~Connolly}

\author[a12]{W.~Cui}

\author[a13]{C.~Duke}

\author[a14]{J.~Dumm}

\author[a15]{A.~Falcone}

\author[a16,a17]{S.~Federici}

\author[a5]{D.~J.~Fegan}

\author[a18]{S.~J.~Fegan}

\author[a12]{J.~P.~Finley}

\author[a19]{G.~Finnegan}

\author[a14]{L.~Fortson}

\author[a12]{J.~Gaidos}

\author[a2]{N.~Galante}

\author[a20]{D.~Gall}

\author[a2]{K.~Gibbs}

\author[a10]{G.~H.~Gillanders}

\author[a21]{S.~Griffin}

\author[a11]{J.~Grube}

\author[a11]{G.~Gyuk}

\author[a21]{D.~Hanna}

\author[a18]{D.~Horan}

\author[a22]{T.~B.~Humensky}

\author[a20]{P.~Kaaret}

\author[a23]{M.~Kertzman}

\author[a5]{Y.~Khassen}

\author[a19]{D.~Kieda}

\author[a7]{H.~Krawczynski}

\author[a9]{F.~Krennrich}

\author[a10]{M.~J.~Lang}

\author[a24,a25]{J.~E.~McEnery}

\author[a9]{A.~S.~Madhavan}

\author[a1]{P.~Moriarty\corref{cor1}}
\ead{pat.moriarty@gmit.ie}
\cortext[cor1]{Principal corresponding author}

\author[a14]{T.~Nelson}

\author[a5]{A.~O'Faol\'{a}in de Bhr\'{o}ithe}

\author[a3]{R.~A.~Ong}
                                                     
\author[a9]{M.~Orr}

\author[a26]{A.~N.~Otte}

\author[a27,a28]{J.~S.~Perkins}

\author[a29]{D.~Petry}

\author[a30]{A.~Pichel}

\author[a16,a17]{M.~Pohl}

\author[a5]{J.~Quinn}

\author[a21]{K.~Ragan}

\author[a31]{P.~T.~Reynolds}

\author[a2]{E.~Roache}

\author[a30]{A.~Rovero}

\author[a2]{M.~Schroedter}

\author[a12]{G.~H.~Sembroski}

\author[a19]{A.~Smith}

\author[a16,a17]{I.~Telezhinsky}

\author[a12]{M.~Theiling}

\author[a10]{J.~Toner\fnref{fn1}} 
\fntext[fn1]{Current address: Valeo Vision Systems, Tuam, Co. Galway, Ireland} 

\author[a21]{J.~Tyler}

\author[a12]{A.~Varlotta}

\author[a32]{M.~Vivier}

\author[a33]{S.~P.~Wakely}

\author[a7]{J.~E.~Ward}

\author[a2]{T.~C.~Weekes\corref{cor2}} 
\ead{weekes@veritas.sao.arizona.edu}
\cortext[cor2]{Corresponding author}

\author[a9]{A.~Weinstein}

\author[a16,a17]{R.~Welsing}

\author[a4]{D.~A.~Williams}

\author[a3]{S.~Wissel}

\address[a1]{School of Science, Galway-Mayo Institute of Technology, Dublin Road, Galway, Ireland}
\address[a2]{Fred Lawrence Whipple Observatory, Harvard-Smithsonian Center for Astrophysics, Amado, AZ 85645, USA}
\address[a3]{Department of Physics and Astronomy, University of California, Los Angeles, CA 90095, USA}
\address[a4]{Santa Cruz Institute for Particle Physics and Department of Physics, University of California, Santa Cruz, CA 95064, USA}
\address[a5]{School of Physics, University College Dublin, Belfield, Dublin 4, Ireland}
\address[a6]{Department of Physics, University of Leeds, Leeds, LS2 9JT, Yorkshire, England, UK}
\address[a7]{Department of Physics, Washington University, St. Louis, MO 63130, USA}
\address[a8]{European Space Astronomy Center, 28080 Villafranca del Castillo, Madrid, Spain}
\address[a9]{Department of Physics and Astronomy, Iowa State University, Ames, IA 50011, USA}
\address[a10]{School of Physics, National University of Ireland Galway, University Road, Galway, Ireland}
\address[a11]{Astronomy Department, Adler Planetarium and Astronomy Museum, Chicago, IL 60605, USA}
\address[a12]{Department of Physics, Purdue University, West Lafayette, IN 47907, USA}
\address[a13]{Department of Physics, Grinnell College, Grinnell, IA 50112-1690, USA}
\address[a14]{School of Physics and Astronomy, University of Minnesota, Minneapolis, MN 55455, USA}
\address[a15]{Department of Astronomy and Astrophysics, 525 Davey Lab, Pennsylvania State University, University Park, PA 16802, USA}
\address[a16]{Institute of Physics and Astronomy, University of Potsdam, 14476 Potsdam, Germany}
\address[a17]{DESY, Platanenallee 6, 15738 Zeuthen, Germany}
\address[a18]{Laboratoire Leprince-Riguet, \'{E}cole Polytechnique, CNRS/IN2P3, Palaiseau, France}
\address[a19]{Department of Physics and Astronomy, University of Utah, Salt Lake City, UT 84112, USA}
\address[a20]{Department of Physics and Astronomy, University of Iowa, Van Allen Hall, Iowa City, IA 52242, USA}
\address[a21]{Physics Department, McGill University, Montreal, QC H3A 2T8, Canada}
\address[a22]{Department of Physics and Astronomy, Barnard College, Columbia University, NY 10027, USA}
\address[a23]{Department of Physics and Astronomy, DePauw University, Greencastle, IN 46135-0037, USA}
\address[a24]{NASA Goddard Space Flight Center, Greenbelt, MD 20771, USA}
\address[a25]{Department of Physics and Department of Astronomy, University of Maryland, College Park, MD 20742, USA}
\address[a26]{School of Physics and Center for Relativistic Astrophysics, Georgia Institute of Technology, 837 State Street NW, Atlanta, GA 30332-0430, USA}
\address[a27]{CRESST and Astroparticle Physics Laboratory NASA/GSFC, Greenbelt, MD 20771, USA}
\address[a28]{University of Maryland, Baltimore County, 1000 Hilltop Circle, Baltimore, MD 21250, USA}
\address[a29]{ALMA Regional Centre, ESO, Karl-Schwarzschild-Strasse 2, 85748 Garching, Germany}
\address[a30]{Instituto de Astronom\'{i}a y F\'{i}sica del Espacio, Casilla de Correo 67 - Sucursal 26 (C1428ZAA), Ciudad Aut\'{o}noma de Buenos Aires, Argentina}
\address[a31]{Department of Applied Physics and Instrumentation, Cork Institute of Technology, Cork, Ireland}
\address[a32]{Department of Physics and Astronomy and the Bartol Research Institute, University of Delaware, Newark, DE 19716, USA}
\address[a33]{Enrico Fermi Institute, University of Chicago, Chicago, IL 60637, USA}

\begin{abstract}
The variability of the blazar Markarian~421 in TeV gamma rays over 
a 14-year time period has been explored with the Whipple 10~m telescope. It is 
shown that the dynamic range of its flux variations is large and 
similar to that in X-rays. A correlation between the X-ray and TeV 
energy bands is observed during some bright flares and when the complete 
data sets are binned on long timescales. The main database consists of 878.4~hours of 
observation with the Whipple telescope, spread over 783 nights.
The peak energy response of the telescope was 400~GeV with 20\% 
uncertainty.
This is the largest database of any TeV-emitting active galactic nucleus (AGN) and hence was used to explore
the variability profile of Markarian~421. 
The time-averaged flux from Markarian~421 over this period was $0.446\pm 0.008$~Crab
flux units. The flux exceeded 10~Crab flux units on three
separate occasions. For the 2000--2001 season the average flux reached 1.86~Crab units, 
while in the 1996--1997 season the average flux was only 0.23~Crab units.

\end{abstract}


\maketitle

\section{Introduction}

Blazars are the most powerful active galactic nuclei (AGN) 
observed and are remarkable for their variability. Their most prominent 
characteristic is the relativistic
jets which are aligned with the line 
of sight to the observer. In this preferred direction the blazar  
can be very bright. Although they are detectable over a wide 
range of frequencies, their alignment does not permit spatial 
resolution and makes the detection of optical spectral lines very 
difficult. Their broadband emission is clearly nonthermal and 
the nature of the progenitor particles (electrons or protons) 
is uncertain. In principle, there is much observational 
evidence which should make their understanding straightforward, 
but in practice there is a wealth of often contradictory 
observations which demonstrate that the underlying mechanisms 
are complex and often ambiguous. While their variability makes these 
AGN inherently interesting, observations 
(preferably simultaneous) over a long time interval at a 
variety of wavelengths are required to draw any firm conclusions. 

The spectral energy distribution (SED) of blazars is generally
represented by a double-peaked structure, with the lower-energy peak 
modeled by a synchrotron emission mechanism in which the 
synchrotron photons arise from electrons in the 
relativistic beam; the Compton scattering 
of these electrons on the low-energy photons (which may be
either of synchrotron origin or external from some
other mechanism) results in high-energy gamma-ray emission 
\citep{ulrich97}. 
Generally, the former simple model (i.e., self-Compton) is preferred, but 
some sources appear to require external photons \citep{dermer93}.
In another class of model, hadronic interactions are invoked
to explain the higher-energy peak \citep{pohl00}. 

Markarian 421 (Mrk421) was one of 70 AGN reported in the 
Third EGRET Catalog of 100~MeV sources \citep{michelson92,hart99}. 
It was remarkable in that it was both the weakest and 
the closest (redshift $z = 0.031$) of these AGN. Its variability was not 
particularly noteworthy in the discovery gamma-ray 
observation. Subsequently it was determined that, 
unlike the other AGN in the EGRET catalog, it should be 
classified as a HBL (high-frequency-peaked BL Lacertae object) in which 
the synchrotron and Compton peaks in its spectral 
energy distribution were displaced some decades 
to higher frequency relative to the norm of EGRET-detected 
AGN. 
This made it a prime candidate for 
TeV emission. It was on a short list of 
EGRET-detected AGN selected for observation 
at TeV energies with the Whipple 10~m telescope \citep{kerrick95}.  
It was the only one on the list which gave evidence of 
a signal, and the subsequent paper \citep{punch92} 
announced the first unambiguous detection of an 
extragalactic source at TeV energies.

Since that time, Mrk421 has been observed every season at 
the Whipple Observatory. Although more than 50 AGN 
have now been detected at TeV energies \citep{wakely12},
Mrk421 is still, on average, the strongest AGN source 
of TeV gamma rays in the northern hemisphere. Because of 
its proximity and high degree of variability, it was 
considered a good candidate for the
detailed study of an AGN that was a TeV gamma-ray emitter. 
During the large flare in 2001, it
was observed to reach a peak emission of 
13 times the level of the Crab Nebula (the brightest 
known steady TeV source) over a four-minute integration time
and to show other rapid variations \citep{krennrich02}.
 
Atmospheric Cherenkov telescopes are particularly 
useful for the study of HBLs such as Mrk421 because 
of their high sensitivity to rapid variations. 
The Whipple telescope has detected flux variability 
with doubling times as short as 15~minutes \citep{gaidos96}; 
there is some correlation between emission at X-ray keV energies 
and gamma-ray TeV energies \citep{fossati08,horan09}. 
Rapid variations in TeV emission have also been observed in other AGN
\citep{albert07, aharonian07}. 
It has been shown that the energy spectrum of Mrk421 hardens 
with increasing intensity \citep{krennrich02}.

In this paper, we summarize intensive observations of Mrk421 
with a single gamma-ray telescope over a long period.
The principal telescope used in this study, the 10~m reflector and imaging
camera at the Whipple Observatory, is described in Section~2.
Section~3 describes the database of TeV observations taken over the 
14-year epoch. Correlations with simultaneous X-ray observations
are presented in Section~4. The level of variability in the TeV signal
is explored in Section~5.  Section~6 explores those 
intervals in which the TeV signal was particularly strong 
and the source may be said to have been flaring. 
 
\section{Whipple Observations.}
The principal data reported here were taken at TeV 
energies with the Whipple 10~m atmospheric Cherenkov 
telescope and imaging camera over a 14-year period 
(December 1995 -- May 2009). Although observations were made 
both before and after this period, this time interval 
was chosen as it represents the best period of uniform 
operation and performance of the telescope.

The Whipple photomultiplier (PMT) camera evolved during 
this time \citep{kildea07}. From 2001 to 2009, the camera 
consisted of a hexagonal array of 379~PMTs of diameter 1.2~cm, giving a total field of
view of $2.8^\circ$. Winston light cones in 
front of the PMTs minimized the light loss and 
gave a relatively uniform sensitivity across the face of the camera. 
The camera was triggered when the light level in at least three tubes exceeded a preset threshold. 
Before 2001, the camera had PMTs of 2.5~cm diameter and had a larger field of view.

The Whipple observations were supplemented by observations with
the VERITAS (Very Energetic Radiation Imaging Telescope Array
System) observatory, located at the basecamp of the
Fred Lawrence Whipple Observatory in southern Arizona, USA;
this was completed in June 2007. The observatory consists of four 12m-diameter imaging
atmospheric Cherenkov telescopes, with a typical baseline 
between telescopes of $\sim$100~m \citep{weekes02}.
Each telescope has a 499-PMT camera, spanning a field of view of $3.5^\circ$.

The Whipple telescope observations 
were taken in intervals of 28 sidereal minutes 
during which the telescope was directed at the source. This 
constituted an ON run. In order to estimate the background, the ON run was 
sometimes followed (or preceded) by an OFF run during 
which the telescope tracked a point 30 minutes later (earlier) in right 
ascension but at the same declination, so that telescope traced out 
the same path in azimuth and elevation as the ON run. Often 
no OFF runs were taken, so that the source was 
tracked continuously (TRACKING mode) and the background
 determined from the known image parameter distributions 
in the absence of a source. 

Observations were taken on clear moonless nights. For 
much of this time, Mrk421 was the prime target, and it was 
observed whenever it was at an elevation greater 
than $55^\circ$. Where possible, one or two ON/OFF pairs 
were also taken on the Crab Nebula each night for calibration purposes. 
The rest of the darktime was used to observe other candidate
sources (mostly AGN) or gamma-ray bursts. 

The Cherenkov images of air showers recorded with the telescope were characterized by 
the moments of the angular distribution of light as 
determined by the output of the PMTs.  The division of the 
data into candidate gamma-ray images and background images 
was made on the basis of selection criteria optimized
by Monte Carlo simulations of gamma-ray and hadron 
shower images and checked (and 
again optimized) on known sources such as the Crab Nebula \citep{reynolds93}. 
On-line analysis was performed at the conclusion of each 
data run so that the state of the source was known promptly 
and the observing program adjusted if necessary. 
Only data taken on clear nights (judged by fluctuations in the    
background trigger rate and by the observers' comments) were used in the final database.

At the beginning of each observing season, the trigger threshold was raised or lowered, if required, to compensate for changes in the telescope and in the camera; these changes included replacement of PMTs, recoating of mirrors, upgrades to trigger electronics, optimisation of mirror-facet alignments, and improvements in telescope pointing corrections. The adjustment to the trigger level from one season to the next was never more than $\pm10$\%, and the threshold was always maintained well above the noise trigger level. 

In order to analyze all the Mrk421 data in a uniform 
manner, the average Crab Nebula flux for each observing season was 
used for calibration. Although the MeV-GeV flux from the 
Crab Nebula has recently been shown to exhibit short-term 
flares \citep{tavani11,abdo11,buehler12}, they are infrequent ($<$2 per year) and their duty cycle is short ($\sim$10~days); a search 
of the Whipple database \citep{anna2012} did not show any
evidence for such flares over a ten-year epoch. Long-term secular variations such as
those seen in X-ray experiments \citep{wilson11} have not been
seen at TeV gamma-ray energies, so that the Crab 
Nebula can still be regarded as a standard candle in this energy regime. 
 
On-source observations of Mrk421 (ON runs and TRACKING runs) 
were chosen and selection criteria as described above applied to identify gamma-ray candidate events. 
Estimation of the background was based on the distribution of the angle $\theta$ between the major axis of the event image and the line from camera centre to image centroid. The region $\theta < 15^{\circ}$ was taken as the signal region and the region $20^{\circ} \leq \theta < 65^{\circ}$ as control region. 
The background in the signal region was estimated by multiplying the number of events in the control region by a ``tracking ratio''. The tracking ratio $\rho$ for each season was determined from a set of ``darkfield'' runs (i.e., observing runs with no known source in the field of view) as the number of size- and shape-selected events in the signal region divided by the corresponding number of events in the control region \citep{catanese98}. Usually the OFF runs from ON/OFF observations of Mrk421 were used as darkfield runs, but where these were insufficient they were supplemented by OFF data on other sources. 
For an on-source run of duration $t$, the gamma-ray rate was then found as $g = (N_{\rm{on}} - \rho N_{\rm{off}})/t$, where $N_{\rm{on}}$ and $N_{\rm{off}}$ are the number of size- and shape-selected events in the signal and control regions, respectively. 
The error on the rate was calculated as $\Delta g = \sqrt{N_{\rm{on}} + \rho^2 N_{\rm{off}} + (\Delta\rho)^2 N_{\rm{off}}^2}$, where $\Delta\rho$ is the error on the tracking ratio (typically $\sim$1--3\%). 
The significance of the detection was taken as $g/\Delta g$; this formulation was used in preference to that of Li~\& Ma \citep{lima83} in order to take account of the error on the tracking ratio. 

Since gamma-ray observations could only be taken with the 
Whipple 10~m telescope on moonless nights, there were about 
eight nights around the time of full moon of each lunar cycle when 
no gamma-ray observations could be made. The periods 
($\sim$21 days each) during which the gamma-ray data were taken are 
referred to as ``darkruns''. Typically, each observing season 
for Mrk421 at the Whipple Observatory spans between 6 
and 8 darkruns, from November--December to May--June.

\section{Gamma-ray Database}

The gamma-ray data for Mrk421 between December 1995 and May 2009, 
in runs of (generally) 28-minute duration, comprise the database for this work; in all cases, the 
measured flux is given in Crab units based on the Crab signal
recorded that season. These Mrk421 runs were combined (weighted average) to give the 
mean flux over larger time intervals: nightly (individual nights 
of 1--6 hour duration), monthly (entire dark runs) and
yearly. The peak energy response of the telescope during these 
observations was 400~GeV with a 20\% uncertainty.

\begin{figure}
\centering {
\includegraphics[scale=0.6]{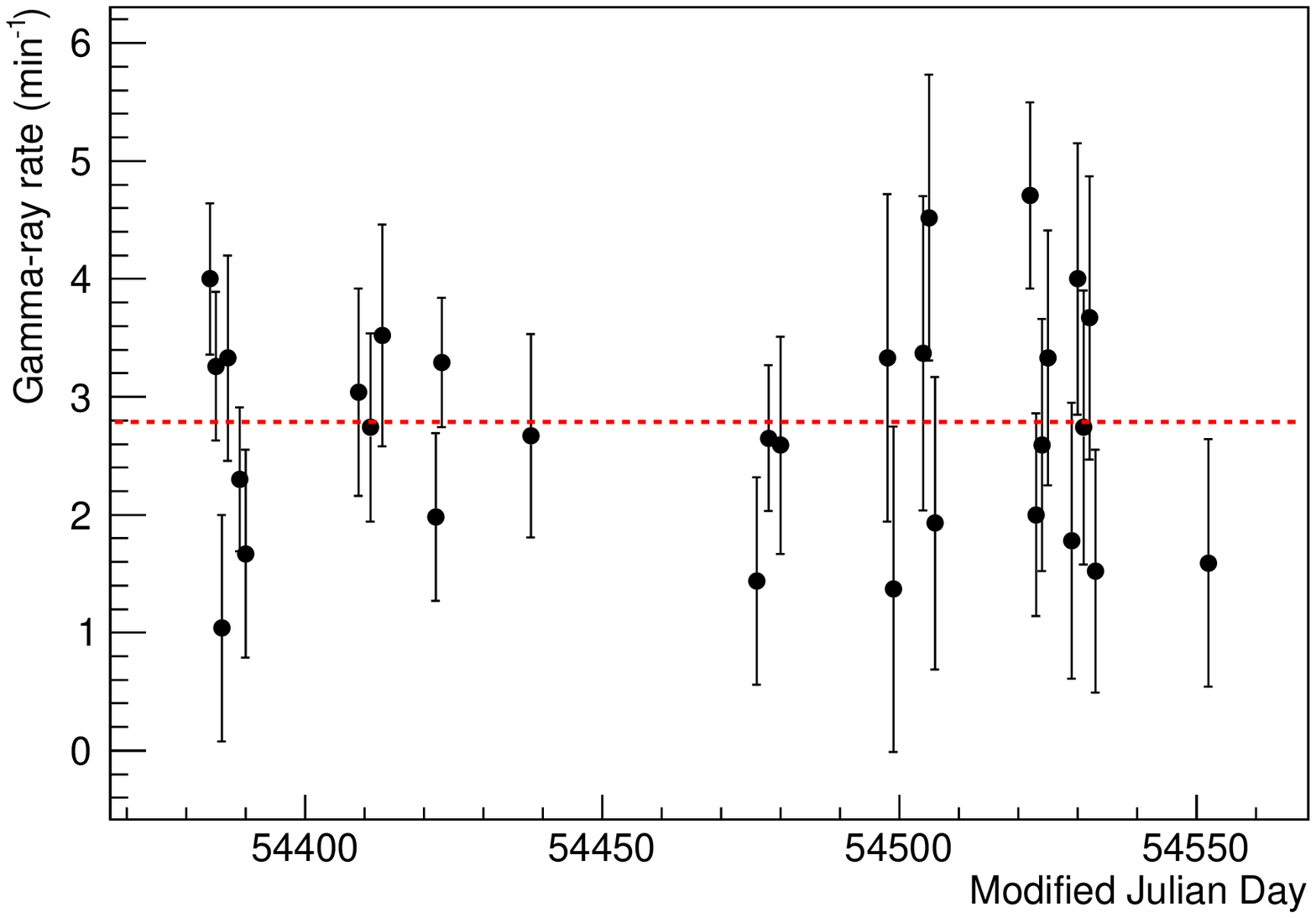}
\caption{Nightly gamma-ray rate from the Crab Nebula from October
2007 to March 2008 at energies $E>$~400~GeV.  The mean rate for the 
season is represented
by the dashed line. The $\chi^{2}$ value is 31.6 for
29 degrees of freedom, giving a probability of 34\%
that the rate is constant.}
\label{fig1}
}
\end{figure}

In this work, the assumption is made that the Crab Nebula 
flux at TeV energies is constant over the season. 
The typical nightly rate from 
the Crab Nebula as measured by the Whipple telescope for one season
is shown in Figure~\ref{fig1}. 
Although the energy spectrum of Mrk421 has been observed to harden 
when the flux level is high \citep{krennrich02}, the spectral shape 
in the energy range 400~GeV to 4~TeV is typically similar to that of the 
Crab Nebula, so that for most nights (when Mrk421 is not flaring strongly) the rates expressed in Crab units 
are insensitive to small changes in peak energy response.
Some results from this data set have been published previously 
\citep{kraw00,blaze05,rebillot06,fossati08,horan09}. 
Several of these reports were motivated 
by the observation of exceptional activity (May 1996 
\citep{gaidos96}; February 2001 \citep{fossati08}; 
May 2008 \citep{ong08}) and the reports included light curves 
on short timescales and spectral analysis. These flares 
are discussed in Section~6. 

\begin{figure}
\centering {
\includegraphics[scale=0.6]{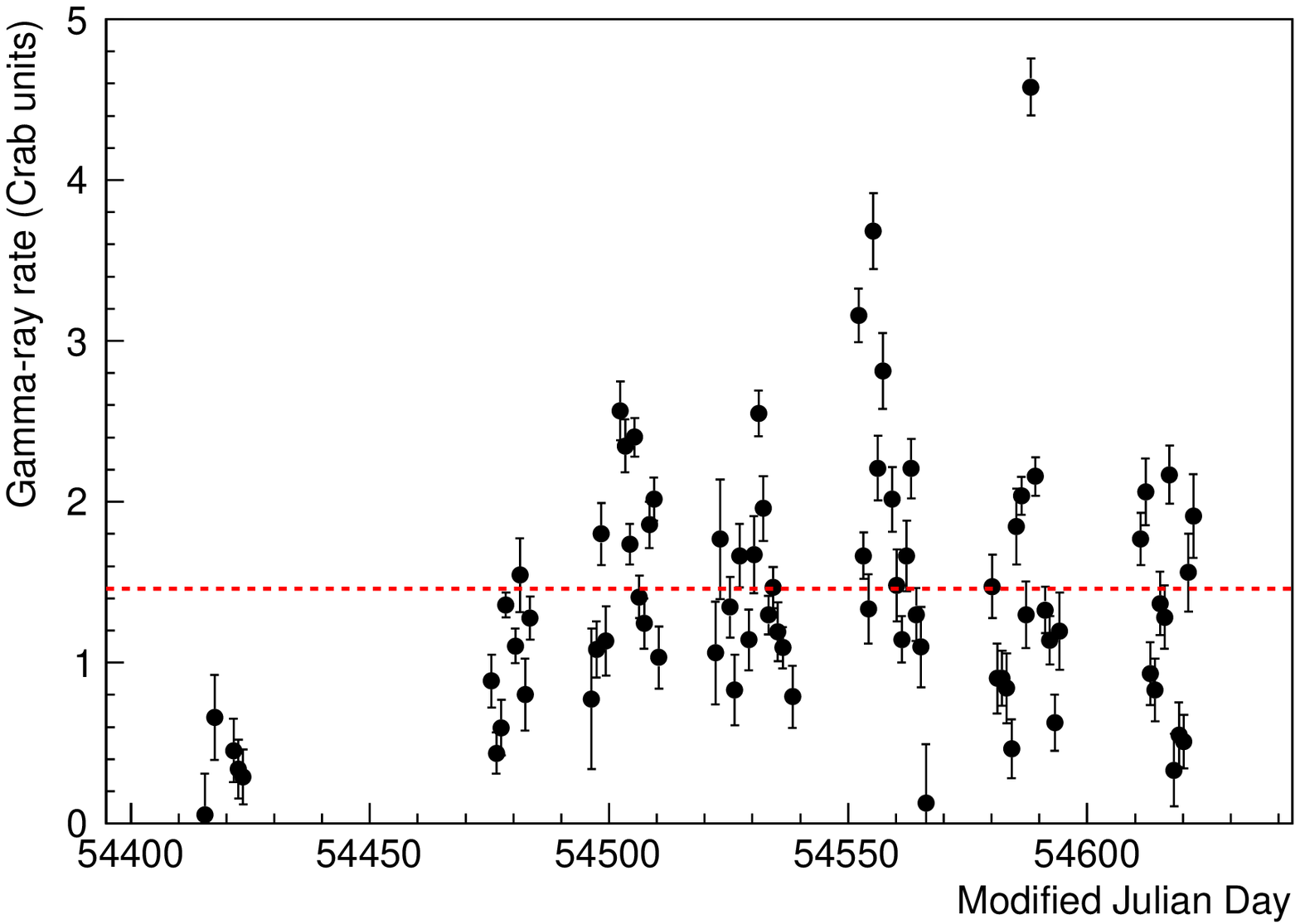}\\
\includegraphics[scale=0.6]{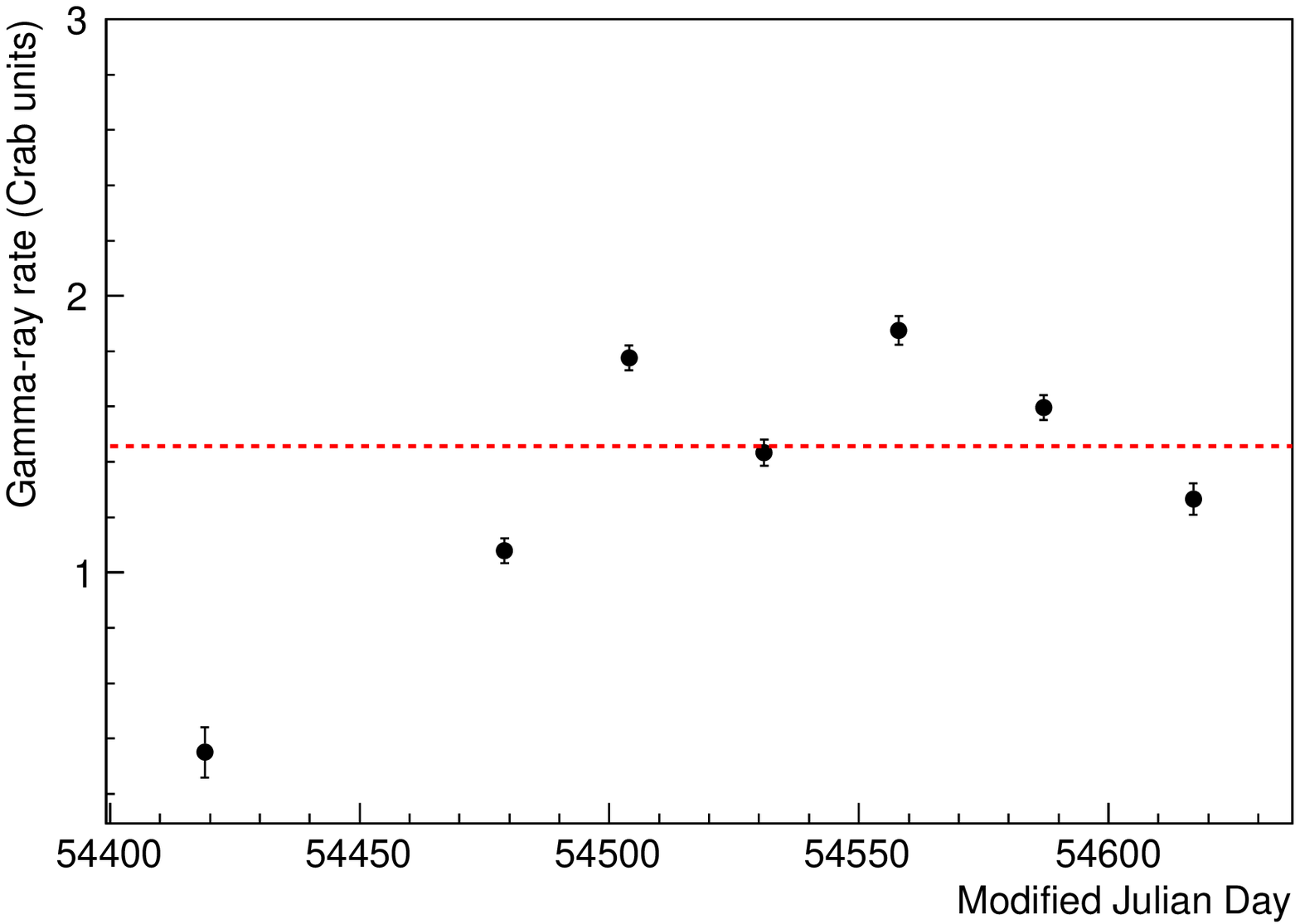}
\caption{The gamma-ray light curve for Mrk421 at energies $E>$~400~GeV 
 for the 2007--2008 
observing season. Top: daily gamma-ray rate; bottom: monthly 
gamma-ray rate. The dashed lines represent the mean rate for the 2007--2008 season.}
\label{fig2}
}
\end{figure}

The Mrk421 observations reported here represent a total 
exposure time of 878.4~hours over 14 years. The nightly gamma-ray 
rate from Mrk421 for the typical 2007--2008 observing season, 
expressed in Crab units, is shown in Figure~\ref{fig2}; the horizontal line  
indicates the average rate for this season.
Mrk421 varies on many time scales.  

\begin{figure}
\centering{ 
\includegraphics[scale=0.6]{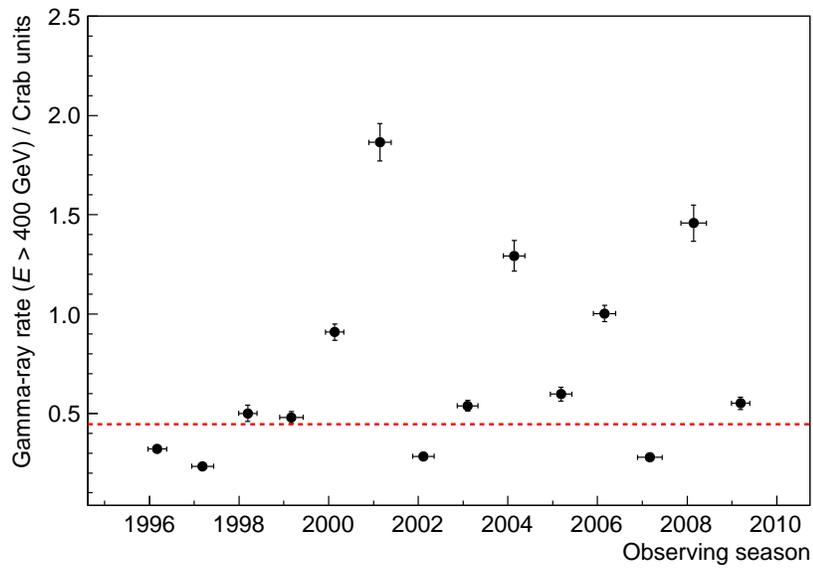}
\caption{The gamma-ray light curve for Mrk421 at energies $E>$~400~GeV
 from 1995 to 
2009, showing the mean rate for each observing season; the horizontal bar on each point indicates the duration of the observing season. The dashed
line represents the mean gamma-ray rate ($0.446 \pm 0.008$ Crab units)   
for the entire 14-year dataset.}
\label{fig3}
}
\end{figure}

The mean gamma-ray rate from Mrk421 was calculated for each 
observing season; the results are given in Table~\ref{yeartable}
and plotted in Figure~\ref{fig3}. The table 
includes the ON-source observing time 
(all the time in which the telescope was pointing at Mrk421)
and the statistical 
significance of the detection for each season.  For the hypothesis 
that the annual rate is a constant, equal to the overall mean 
rate for the complete 14-year data set, the $\chi^{2}$ value is 1022
for 13 degrees of freedom, corresponding to a chance probability of 
$P < 10^{-300}$.
During some observing seasons Mrk421
was the brightest TeV source in the sky, whereas              
in others it was barely detectable with the Whipple telescope.

\begin{table}
\centering
\caption{\normalsize{Yearly Summary of Mrk421 observations}}
\begin{tabular}{c c c c}

\hline

Season          &   Exposure       & Significance & Mean gamma-ray rate\\
         & (hours)  &  (standard deviation) &    (Crab units)\\   

\hline
1995-1996 &  \ \ 51.8   &    36  &  0.32 $\pm$ 0.02      \\
1996-1997 &  \ \ 53.2   &    25  &  0.23 $\pm$ 0.02      \\
1997-1998 &  \ \ 52.1   &    24  &  0.50 $\pm$ 0.04      \\
1998-1999 &  \ \ 45.2   &    23  &  0.48 $\pm$ 0.03      \\
1999-2000 &  \ \ 38.7   &    36  &  0.91 $\pm$ 0.04      \\
2000-2001 &  \ \ 78.1   &    88  &  1.86 $\pm$ 0.09      \\
2001-2002 &  \ \ 32.9   &    14  &  0.28 $\pm$ 0.02      \\
2002-2003 &  \ \ 45.0   &    35  &  0.54 $\pm$ 0.03      \\
2003-2004 &  \ \ 51.8   &    82  &  1.29 $\pm$ 0.08      \\
2004-2005 &  \ \ 23.0   &    25  &  0.60 $\pm$ 0.04      \\
2005-2006 &  \ \ 75.1   &    88  &  1.00 $\pm$ 0.04      \\
2006-2007 &  \ \ 52.0   &    18  &  0.28 $\pm$ 0.02      \\
2007-2008 &     148.9   &    84  &  1.46 $\pm$ 0.09      \\
2008-2009 &     130.6   &    39  &  0.55 $\pm$ 0.03      \\
\hline
\end{tabular}
\label{yeartable}
\end{table}

\begin{landscape}
\begin{figure}
\centering {
 \includegraphics[scale=0.9]{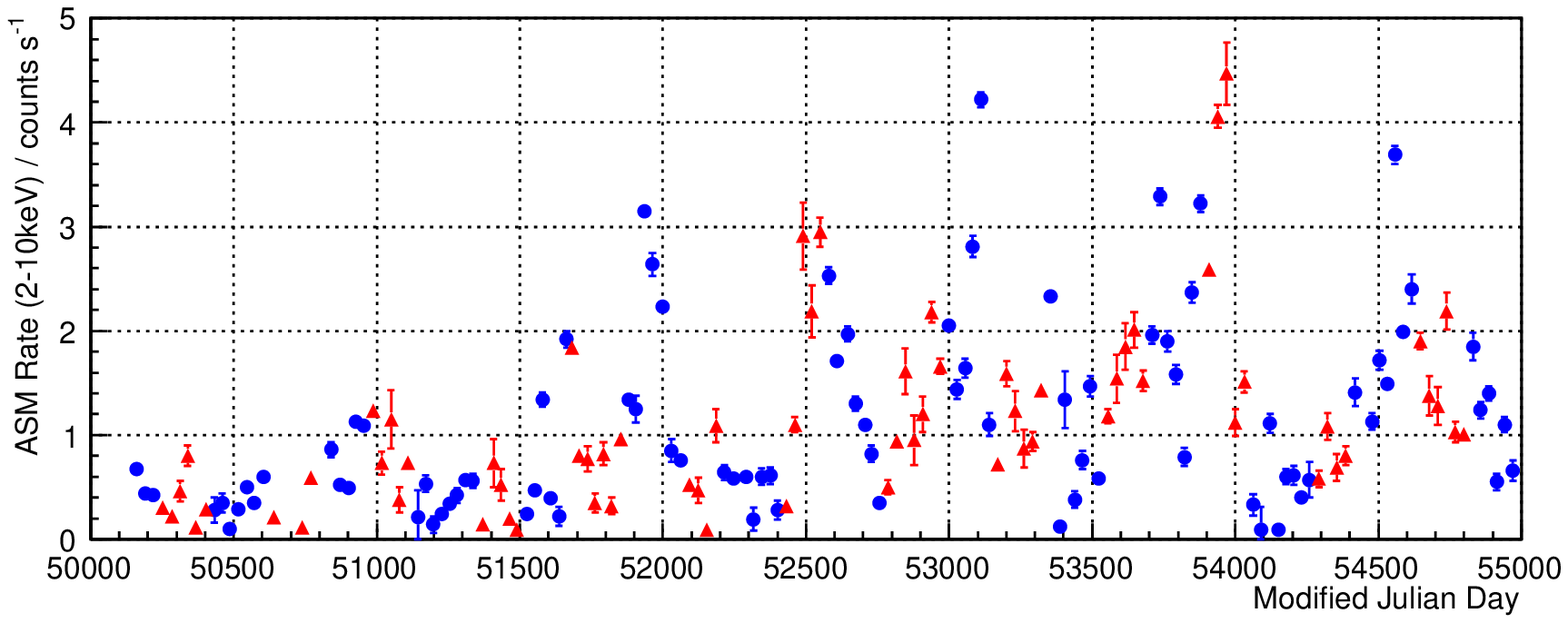} \\
 \includegraphics[scale=0.9]{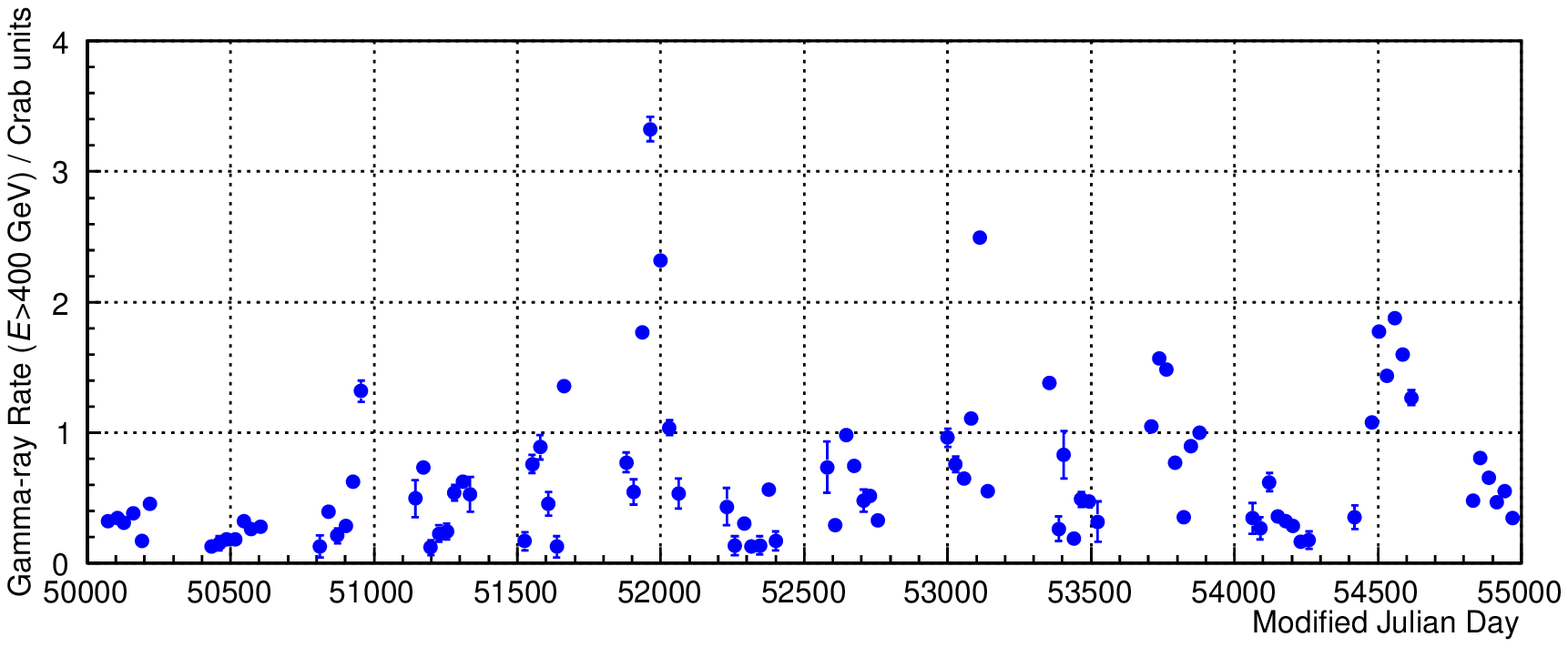}
 \caption{The monthly X-ray and gamma-ray light curves for Mrk421
(1995--2009). Top:  RXTE ASM (2--10 keV); bottom: 
Whipple 10~m telescope ($>$~400~GeV).
The RXTE data are shown as circles for periods during which Mrk421 was observable with the Whipple 10~m telescope (typically from November-December to May-June each season) and as triangles for intervals when it was not.}
\label{fig4}
}
\end{figure}
\end{landscape}

Figure~\ref{fig4} (bottom) shows the gamma-ray rate averaged over each darkrun 
(monthly) for the 14 years of data. With this unique dataset, 
it is possible to look for long-term 
temporal changes in flux levels, slow periodicities and correlation 
with experiments at other wavelengths with long time constants.

\begin{landscape}
\begin{figure}
\centering {
\includegraphics[scale=0.9]{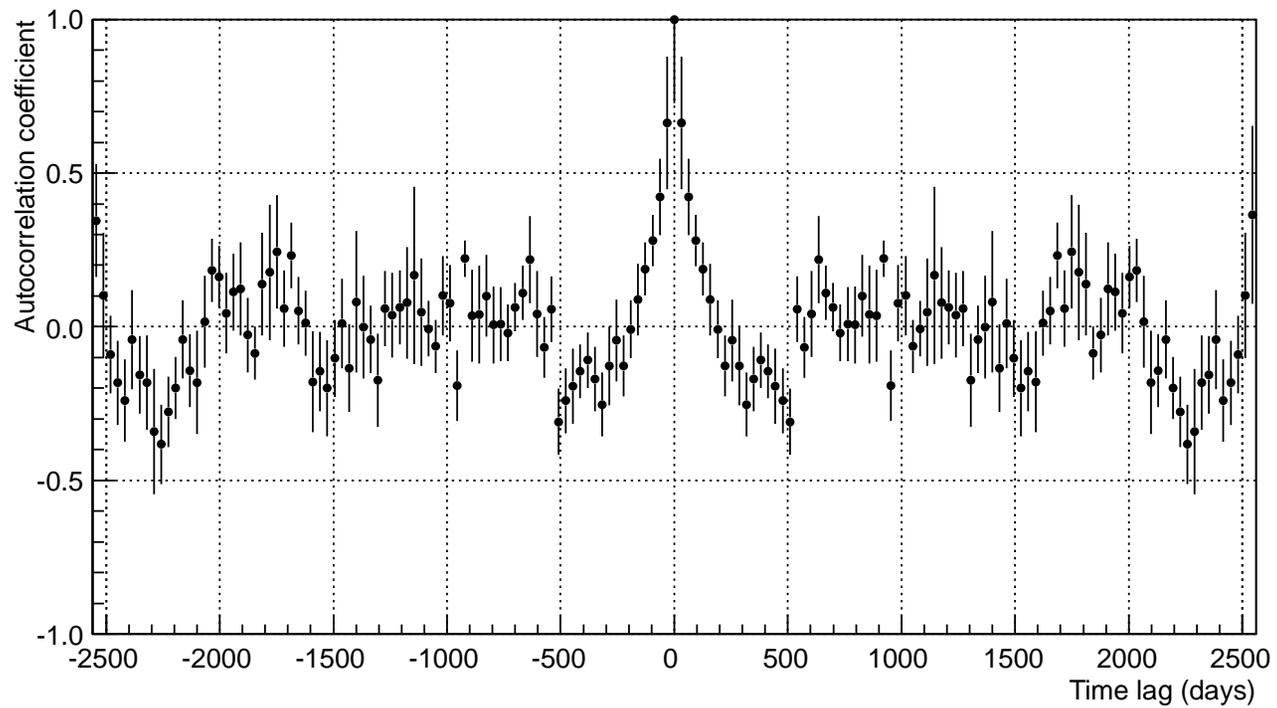}
\caption{The discrete autocorrelation function for the 
monthly-binned Mrk421 Whipple gamma-ray database.}
\label{fig5}
}
\end{figure}
\end{landscape}

The discrete autocorrelation function (DACF) allows the study 
of the level of autocorrelation in unevenly sampled datasets 
\citep{edelson08,hufnagel92} without any interpolation or 
addition of artificial data points. It can reveal the 
presence of periodicity (intrinsic or spurious) in the time
series. The DACF for the gamma-ray dataset for Mrk421 is 
plotted in Figure~\ref{fig5}; it shows no evidence for periodicity 
at any period between 2 months and 7 years. 

\section{Correlations}

We correlated these Whipple 10~m observations with archival 
data taken at other wavelengths for part  or all of the epoch 1995--2009. 
The most complete overlap was with the Rossi X-ray Timing Explorer
(RXTE) database for the entire 14 years \citep{bradt93,levine96}. There 
was also significant overlap with Milagro \citep{smith01}.
There was partial overlap with observations by
Fermi-LAT \citep{fermi} and with data from three radio 
AGN monitoring programs (Mets\"{a}hovi Radio Observatory 
\citep{terasranta98}, the University of Michigan Radio 
Observatory \citep{aller85}, and MOJAVE \citep{lister05}). 
The correlation with the X-ray data from RXTE will be considered
here; the correlation with Milagro, Fermi-LAT and the radio programs
will be the subject of a separate publication.

RXTE has three non-imaging X-ray detectors, one of which is the 
All-Sky Monitor (ASM); it was launched in December 1995 and was 
operational for the period of the Whipple 10~m monitoring program
described here. The ASM scanned nearly 80\% of the sky in each orbit
and was sensitive to X-rays of energy 2--10 keV. Systematic uncertainties
in the ASM rate were large, but relative flux estimates were
available for month-scale monitoring. 

In the ASM database \citep{asm}, each data point represents the one-day average 
of the fitted source fluxes from a number (typically 5--10) of 
individual ASM dwells (90 seconds each). The selection criteria for the observations are described in \citep{asm}. The data are quoted as nominal 2--10 keV rates in ASM counts per second, where the 
Crab Nebula rate is about 75 counts per second. The daily rates were combined (weighted average) in $\sim$21-day intervals to reduce statistical errors.

The gamma-ray rates for the 1995--2009 Mrk421 Whipple observations were 
compared with the X-ray flux from the ASM. 
The light curves for the X-ray and gamma-ray data on Mrk421
from December 1995 to May 2009 are plotted in Figure~\ref{fig4}
(top and bottom, respectively). 
The data are plotted for the Whipple 10~m darkrun periods ($6-8$ for each season) in which both X-ray and 
gamma-ray data were taken, i.e., the X-ray data were averaged 
for each period in which Whipple gamma-ray data were available 
(usually 21-day periods). 

\begin{figure}
\centering {
\includegraphics[scale=0.4]{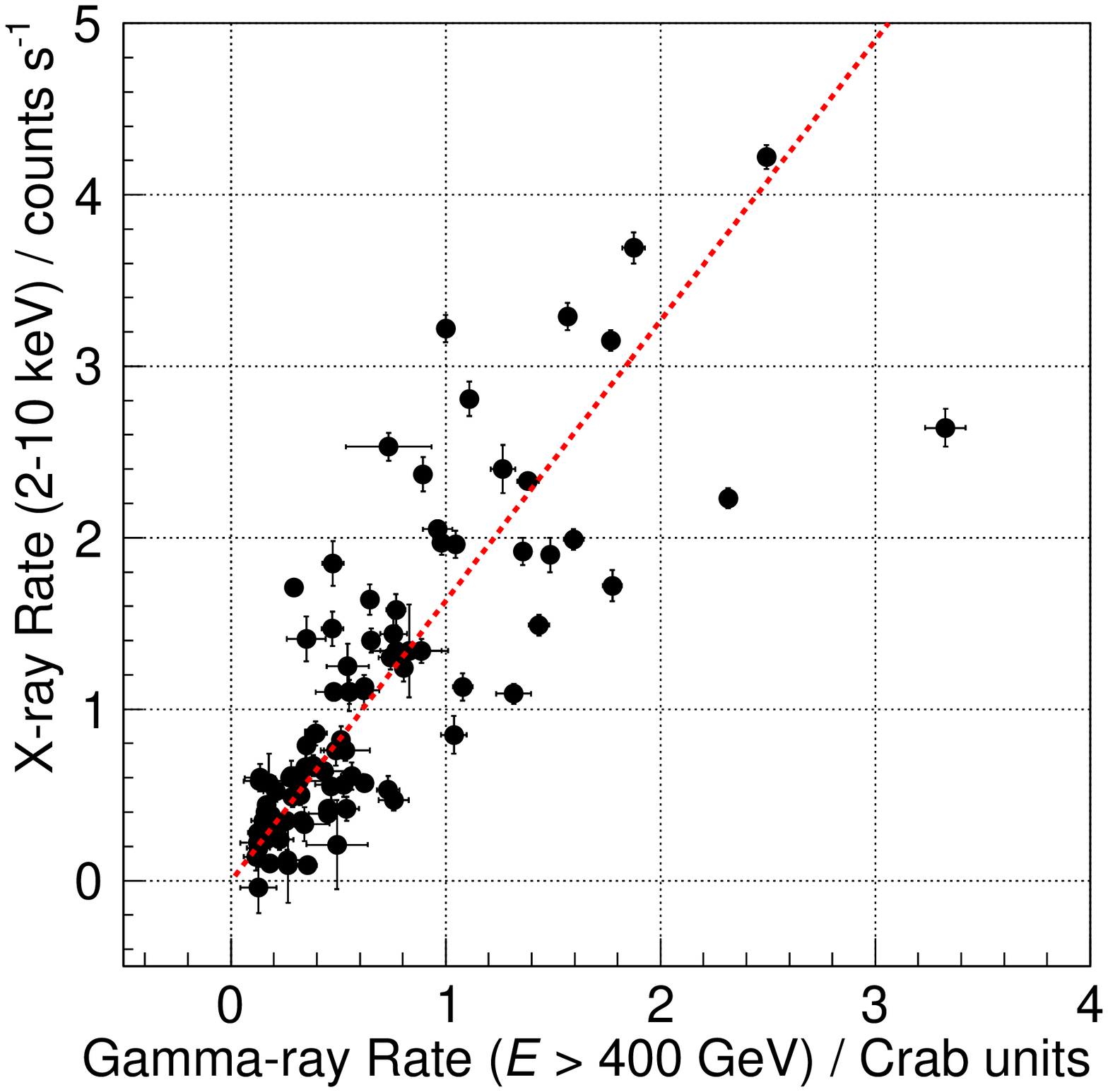} \includegraphics[scale=0.4]{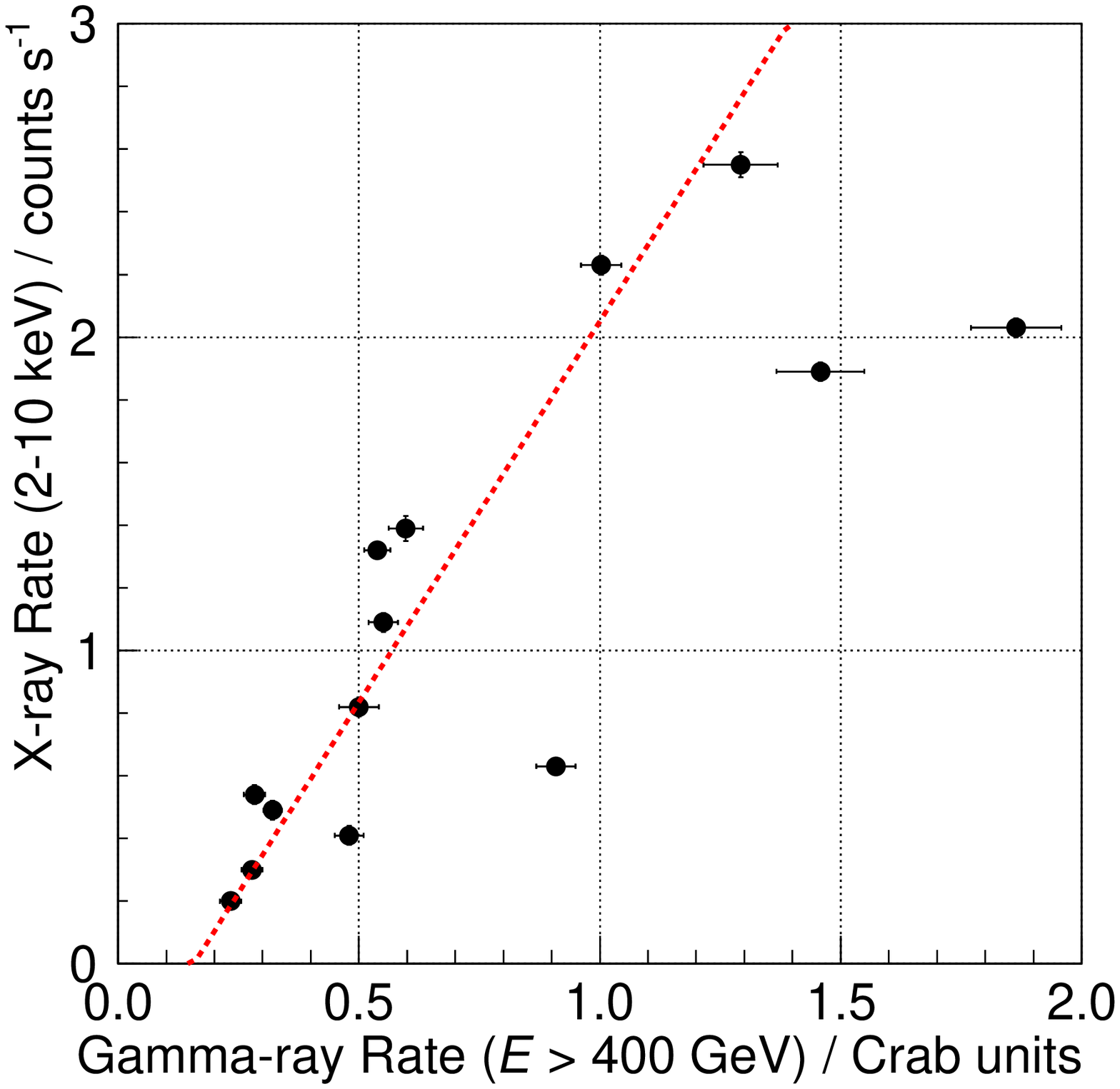}
\caption{Flux-flux correlation for the Whipple 10~m ($>$~400~GeV)
 and RXTE
ASM (2--10 keV) Mrk421 data on two different timescales. The dashed lines are
the best-fit straight lines. Top: monthly-binned gamma-ray
and RXTE ASM data ((the best-fit line has slope 1.63 and correlation coefficient $R = 0.75$)); bottom: annual-binned gamma-ray
and RXTE ASM data (slope 2.43, $R = 0.89$).}
\label{fig6}
}
\end{figure}

\begin{landscape}
\begin{figure}
\centering {
\includegraphics[scale=0.9]{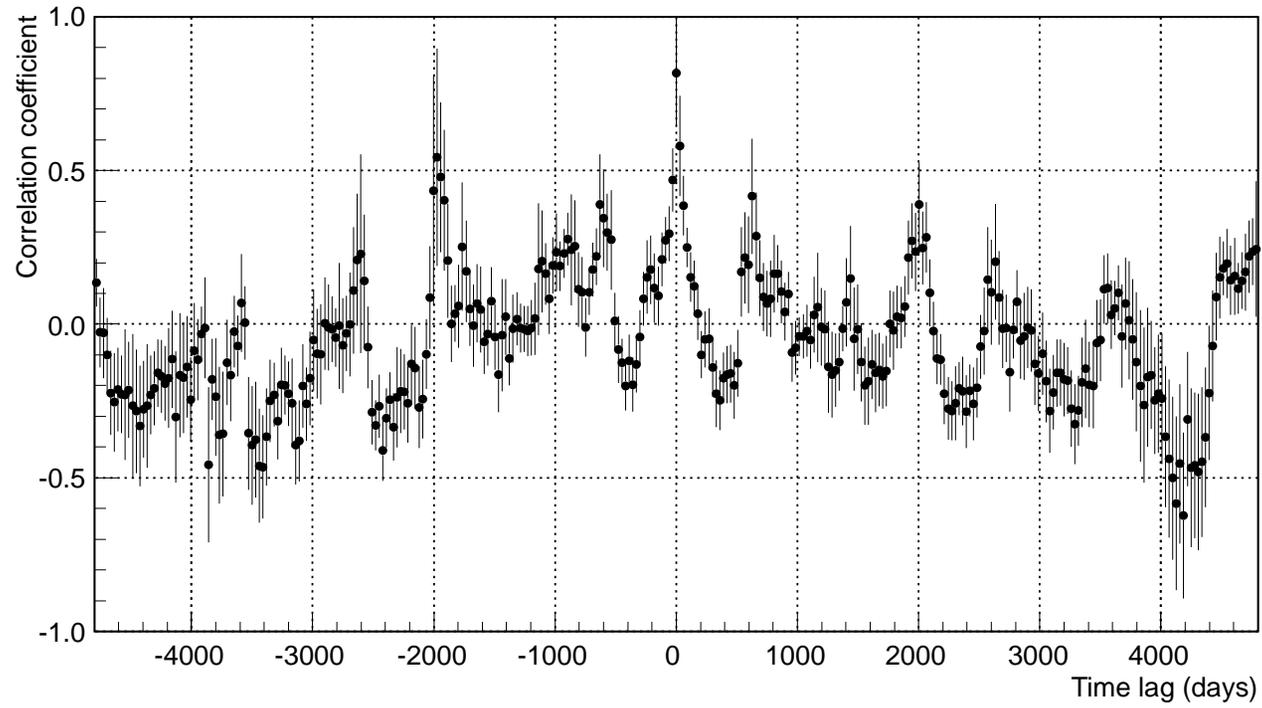}
\caption {The discrete correlation function
between Mrk421 X-ray (2--20 keV) and gamma-ray ($>$~400~GeV) light curves.}
\label{fig7}
}
\end{figure}
\end{landscape}

The RXTE ASM data (1995--2009) show evidence for emission correlated with the 
gamma-ray data on monthly and yearly timescales, as shown 
in Figure~\ref{fig6}. In previous papers \citep{fossati08,horan09,acciari11}, 
the correlation on short timescales was treated in detail. Here, we
concentrate on the longer timescales which are uniquely available 
in our large database.
The correlation coefficients are $R = 0.75$ and 
$R = 0.89$ for the monthly- and yearly-binned data, respectively.
There is general agreement between these values and those obtained
in previous studies of parts of this 
database \citep{fossati08,horan09}. The strong correlation suggests
that the X-rays and gamma rays are emitted by the same population 
of electrons in the same region of the jet. 
During one flaring episode, a $2.1 \pm 0.7$ ks time lag of the TeV flare with respect to the 
2--4 keV X-ray band was observed \citep{fossati08}.  
For the 14-year data set, the discrete correlation function (DCF) 
between the gamma-ray and X-ray monthly-averaged fluxes 
for time lags (gamma ray relative to X-ray) between $-4800$ and $+4800$ 
days is shown in Figure~\ref{fig7}. 
The TeV and keV fluxes are seen to be significantly correlated 
(with a DCF value of 0.82) for a time shift shorter than 30 days; 
the maximum of the DCF is centered at a lag of ($0\pm15$) days.

\section{Variability}

Mrk421 is the brightest AGN at TeV energies (on average) and was 
the first such AGN to be discovered at TeV energies. 
Mrk421 is thus the ideal candidate for the detailed study of the variability
of this type of object, with the caveat that it may have a 
greater degree of variability than the norm for such objects on all 
timescales. It is remarkable also in being the closest TeV-emitting AGN. 
It is certainly the most studied of the known TeV blazars. 
Most of the variations seen in Mrk421 on the various timescales have been 
seen also in other AGN, but the statistics are sparse. It is probably 
the youngest such AGN. Because it is 
so close, it has the 
lowest spectral distortion due to absorption by pair 
production with the extragalactic background light. 

It is difficult to compile a large data sample from a 
variety of observatories, because 
observing modes may be quite different and the relative sensitivity 
of the experiments is 
often not well-defined. The Whipple 10~m database is comparatively 
uniform and covers a long time interval; during 
the period 1995--2009, the system was operated in a relatively 
stable configuration of telescope and camera \citep{kildea07}. 
It is thus useful to use this database to investigate the statistical 
distribution of the time variability of Mrk421 and, by inference, 
that of other AGN. If Mrk421 is typical of all (or a subclass of) 
TeV AGN, then we can use these 
distributions to predict how often a more distant AGN might 
flare or be above a certain level of brightness.  
Four different timescales are considered: yearly, monthly, 
daily and run-by-run (half-hourly). 

\begin{figure}
\centering{
\includegraphics[scale=0.6]{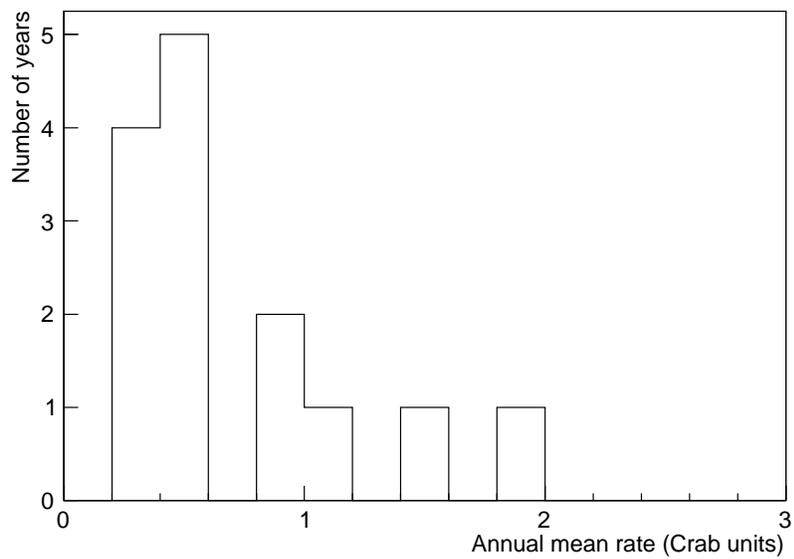}
\caption{Distribution of the yearly-binned Mrk421 gamma-ray rate ($E>$~400~GeV) for 14 years of data (1995--2009).}
\label{fig8}
}
\end{figure}

Yearly variations: As seen in Figure~\ref{fig8}, there are 
significant variations on a yearly timescale, with mean annual 
rates ranging from 0.23 Crab units to 1.86 Crab units. Mrk421 was 
brighter than any other TeV source in the sky during most of the 
2007--2008 observing season (see Figure~\ref{fig2}); this was also the case for the 2000--2001 season.

\begin{figure}
\centering{
\includegraphics[scale=0.6]{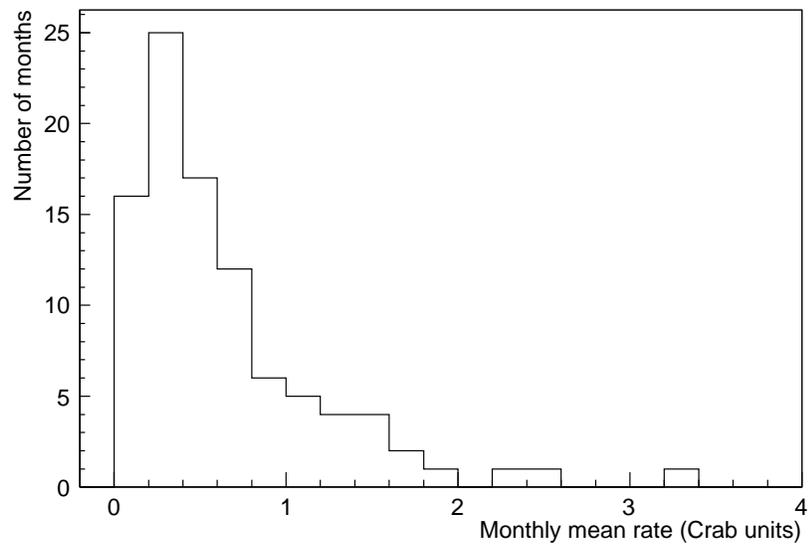}
\caption{Distribution of the monthly-binned Mrk421  
gamma-ray rate ($E>$~400~GeV) for 95 months (darkruns), representing 14 years of data (1995--2009).}
\label{fig9}
}
\end{figure}

Monthly variations (Figure~\ref{fig9}):  There are clearly large 
variations on this timescale over the full 14 years 
of data. There were 19 months (out of a total of 95 months) 
in which the average rate was at least 1 Crab unit.

\begin{figure}
\centering{
\includegraphics[scale=0.6]{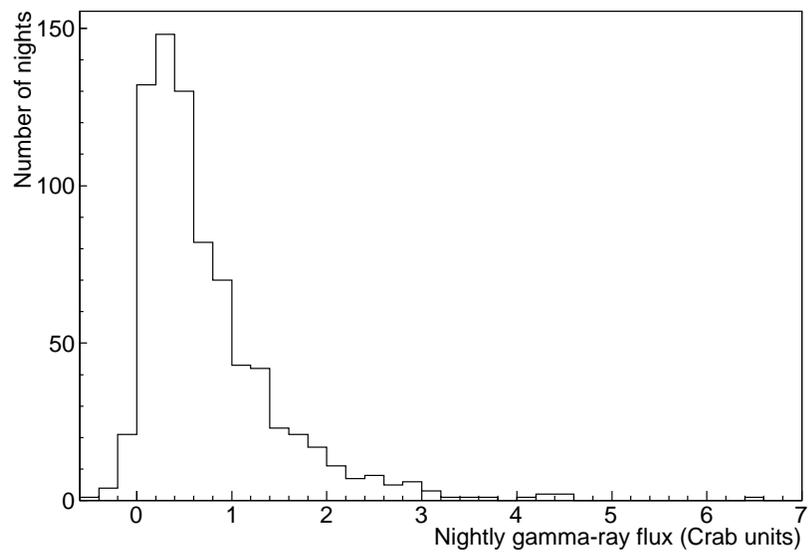}
\caption{Distribution of the nightly-binned Mrk421  
gamma-ray rate ($E>$~400~GeV) for 14 years of data (1995--2009), covering 783 nights.}
\label{fig10}
}
\end{figure}

Nightly variations: The distribution of the average rate for 
each of the 783 nights of observation is 
shown in Figure~\ref{fig10}.  The 
duration of the observations on which the average is based 
varied from night to night as observing 
priorities, weather, etc., changed. There were 49 nights 
(out of a total of 783 nights, 6.3\%) for 
which the average rate was at least 2 Crab units.

\begin{figure}
\centering{
\includegraphics[scale=0.6]{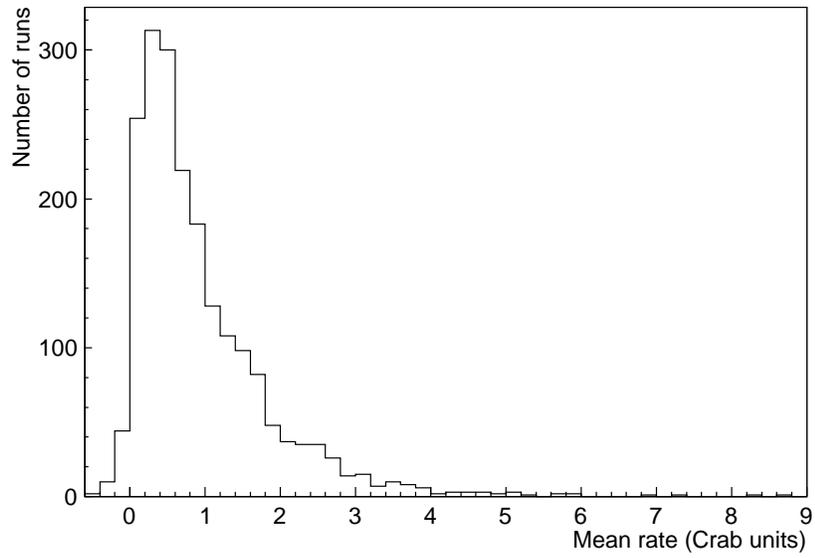}
\caption{Distribution of the Mrk421 
 gamma-ray rate ($E>$~400~GeV) for 2007 individual runs (generally of 28-minute duration) taken over 14 years of observation (1995--2009).}
 \label{fig11}
}
\end{figure}

\begin{landscape}
\begin{figure}
\centering{ 
\includegraphics[scale=0.9]{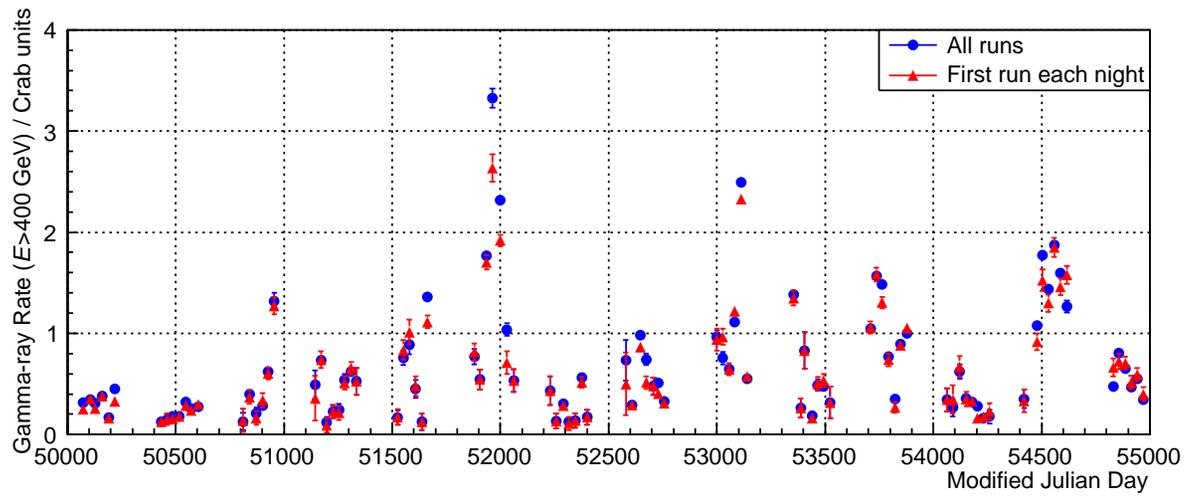}
\caption{Mrk421 gamma-ray flux ($E>$~400~GeV) for each darkrun from 1995 
to 2009. Fluxes calculated using all available data runs (typical 
duration 28 minutes) satisfying weather, hardware and elevation criteria are shown as circles (blue); triangles (red) are fluxes calculated using only the first data run on 
each night of observation.}
\label{fig12}
}
\end{figure}
\end{landscape}

Run variations: Observing runs were generally of 28-minute duration. 
The distribution of the run-by-run rates is shown in 
Figure~\ref{fig11}. This distribution is not a completely 
unbiased sample, since for a subset of the data
there was a tendency for the observers to continue observations when 
the initial observed signal was high.   
There is therefore some bias towards including observations when Mrk421
was in a high state (there is no bias against runs
where the signal was low, as all observations were included in the
analysis irrespective of the apparent flux level). 
However, when only the first run in each night
is considered, this bias was not found to be strong, as can be seen in
Figure~\ref{fig12} which shows the flux using all the data satisfying weather, hardware
and elevation criteria and a smaller set of data where the fluxes are based
on only the first run of each night. In later years, this bias was largely 
removed as observations were concentrated on just Mrk421.
34\% of the runs, taken at random, 
show Mrk421 to be at an emission level of $>1$ Crab unit, and 3.5\% of 
the runs have levels above 3 Crab units. 
The highest fluxes ($>7$ Crab units) were observed in sub-run intervals of 2--5 minutes
(see next section).

If Mrk421 is typical (which is probably not the case), these distributions 
could be used to predict the expected level of variation in other 
TeV-emitting AGN, scaling by 
the ratio of their average TeV gamma-ray intensity to that of Mrk421 
(e.g., for a TeV-emitting AGN one-third as bright as 
Mrk421 on average, $\sim$3.5\% of the runs taken on it might be expected to 
exhibit flaring above the 1 Crab level). 

\section{Large Flares}

Mrk421 is characterized by significant flux variations 
on timescales of a few minutes to years and perhaps longer.
During the 14 years of monitoring reported here, there
were three instances in which the variations were sufficiently
strong that they were regarded as ``flares'' and were so 
reported in subsequent publications \citep{gaidos96,fossati08,acciari11}.
Although the timescales were quite different, the common
factor was that in each case the observed flux reached
a flux level in excess of 10 Crab in a bin of $\leq 5$ minutes.
For completeness, the salient features of these three instances
of exceptional variability are summarized here.

The particular value of these observations was that the
signal-to-noise ratio (number of gamma-ray events to 
background events) was sufficiently large that the 
spectrum of gamma rays could 
be determined. In the normal operation of observing a
relatively weak source with a single Cherenkov telescope, a
parallel series of observations of an off-source
region must be taken before a reliable spectrum can be derived.
Hence, in this long-term monitoring program, where the 
emphasis was on determining the lightcurve and off-source measurements
were not routinely made, the
spectrum could not be reliably derived. 

\begin{figure}
\centering{ 
\includegraphics[scale=0.6]{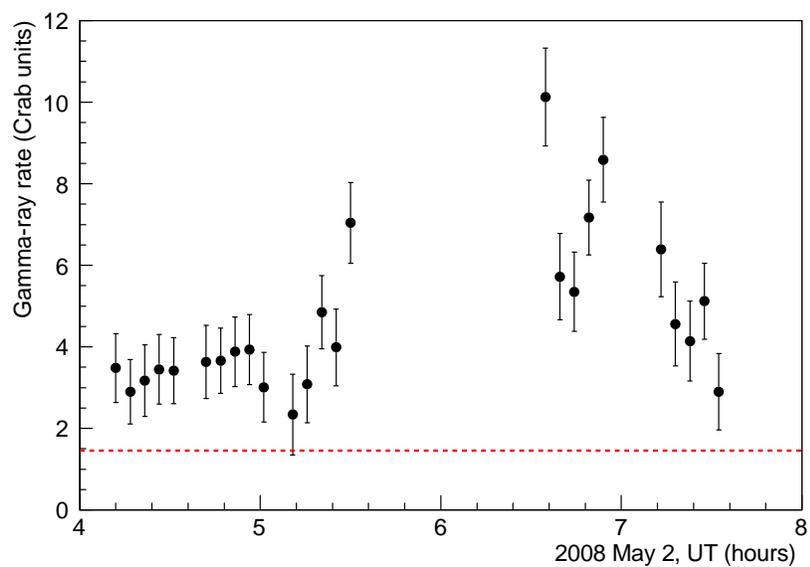}
\caption{Whipple 10~m gamma-ray ($E>$~400~GeV) light curve  
for Mrk421 on 2nd~May~2008 (MJD 54588), binned in 5-minute intervals. The dashed line 
shows the mean rate  
for the 2007--2008 season. The gap in the plot arises from two 
factors: firstly, the data run immediately before the gap was 
taken in ON/OFF mode, and secondly, the telescope had already 
been slewed to observe another target before the analysis 
of the Mrk421 run showed that a major flare was taking place.}
\label{fig13}
}
\end{figure}

(i) During the 1995--1996 season, Mrk421 was relatively
quiet; the mean gamma-ray rate
for the season was $0.32\pm0.01$ Crab, with a significance
of $36\sigma$ for a total exposure of 51.8~hours (Figure~\ref{fig4}).
There were two remarkable outbursts of TeV gamma rays
just eight days apart \citep{gaidos96}.
The first of these occurred on 7th~May~1996 (MJD 50210), when over the course
of 2.5~hours the flux increased
by a factor of five and reached a maximum rate of $\sim$10 Crab
($\sim$30 times the average flux for that season). The observations
had to be terminated while the flux was still rising due to increasing
moonlight so that the peak of the flare was probably not
observed. However, on the following night the flux was down to
$0.39\pm0.05$ Crab, just above the season average. A second flare
of lower intensity, but unusual for its very short duration,
occurred on 15th~May~1996 (MJD 50218), when a rise and fall within $\sim$30 minutes
was observed. The spectrum of the large flare was reported in
\citep{zweerink97} as 
$F(E) = (2.24\pm0.12\pm0.7) E^{-2.56\pm0.07\pm0.1}$~s$^{-1}$~cm$^{-2}$~TeV$^{-1}$,
where the first error is statistical and the
second systematic. This spectrum was confirmed in  
\citep{krennrich99}, where slight evidence for curvature was found. 

ii) In 2007--2008, the Whipple monitoring program again found Mrk421
in a strong flaring state; on 2nd~May~2008 (MJD 54588), the flux 
exceeded 10~Crab \citep{atel1506} in a 5-minute interval (Figure~\ref{fig13}). It
appeared that the observation was during the declining phase
of a flare whose maximum was not observed.
Observations with the VERITAS telescope on the
following night confirmed the high level of gamma-ray activity \citep{acciari11}.

\begin{figure}
 \centering {
\includegraphics[scale=0.6]{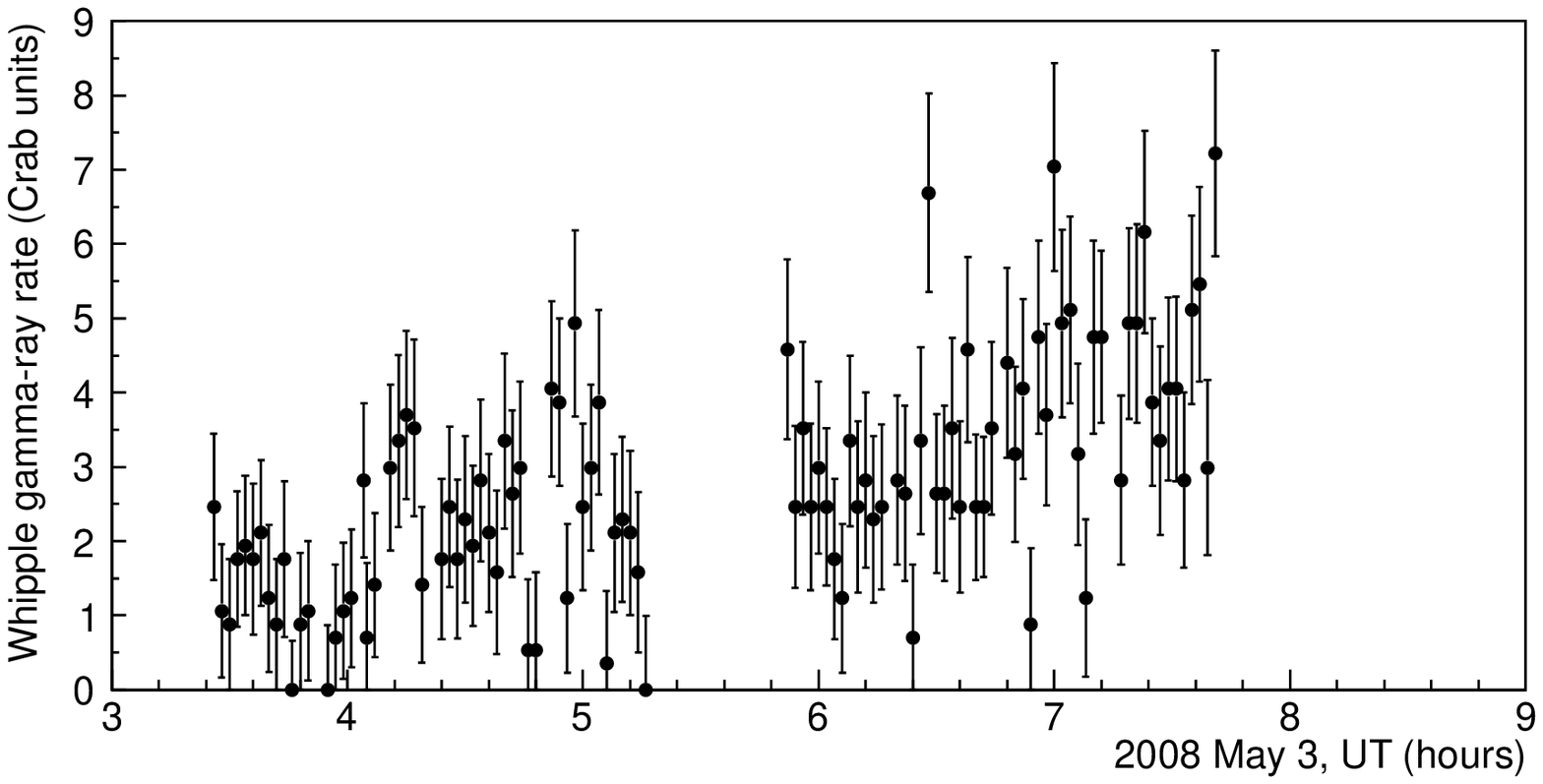} \\
\includegraphics[scale=0.6]{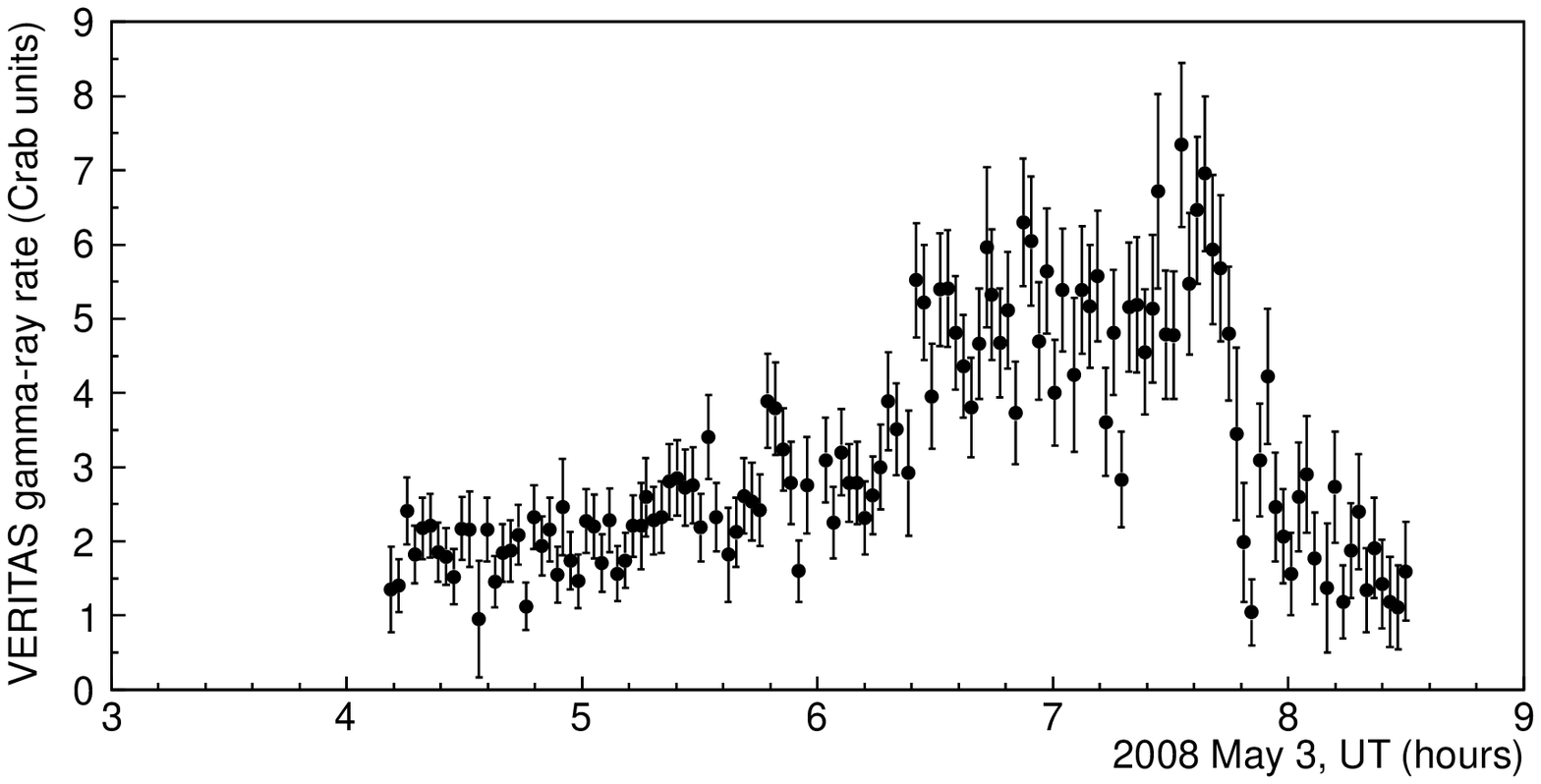} \\
\includegraphics[scale=0.36]{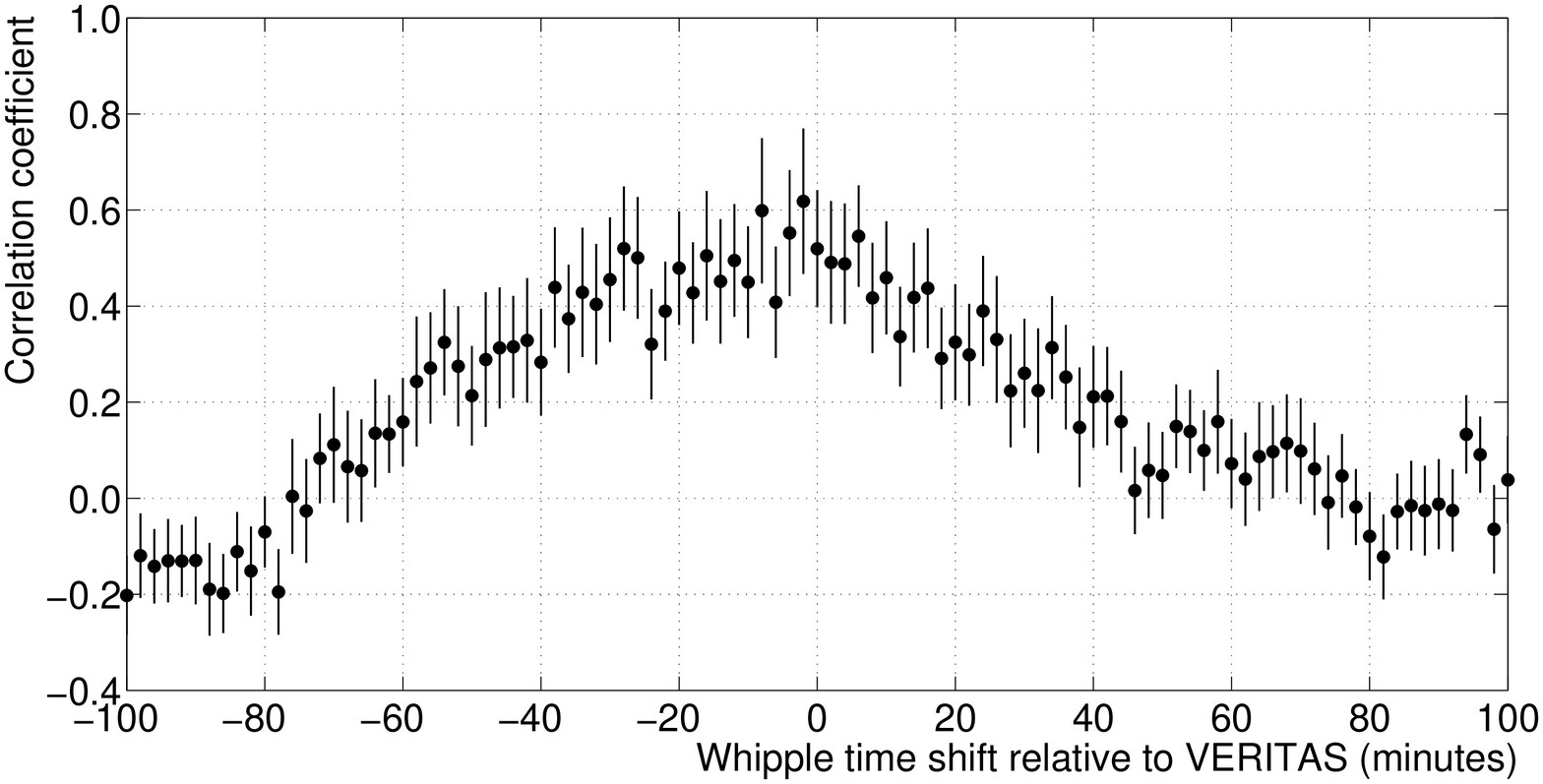}
\caption{Mrk421 gamma-ray rate on 3rd~May~2008 (MJD 54589), from Whipple ($E>$~400~GeV)
and VERITAS ($E>$~300~GeV) observations (top and middle panels, respectively). 
Both light curves are binned in 2-minute intervals. The bottom panel 
shows the DCF between the Whipple and VERITAS data.}
\label{fig14}
 }
\end{figure}

On 3rd~May (MJD 54589), the Whipple telescope and VERITAS observed
Mrk421 \citep{acciari11}, and the light curves are shown in Figure~\ref{fig14}, 
 binned in 2-minute intervals. This is the only
instance in which observations  of the same target with independent imaging
atmospheric Cherenkov systems were carried out simultaneously with
sufficiently high signal-to-noise ratio to permit a meaningful comparison of the
gamma-ray rates. The DCF between the Whipple (single 10~m telescope with
energy threshold $\sim$~400~GeV) and
VERITAS (the array of four 12~m telescopes with energy threshold $\sim$300
GeV) rates, shown in Figure~\ref{fig14}c,
is an interesting
confirmation of the validity of the atmospheric Cherenkov technique.
However, because the observations were at large zenith angles, it was
difficult to derive a reliable spectrum from the Whipple telescope observations.
For the ``very high state'', the spectrum measured with VERITAS was fitted using a power law with exponential cutoff model \citep{krennrich02,aharonian05}, of the form 
$$F(E) = I_0 \left(\frac{E}{E_0}\right)^{-\alpha} \exp\left(-\frac{E}{E_c}\right),$$
\noindent yielding normalization constant $I_0 = (32.0 \pm1.2) \times 10^{-7}$ s$^{-1}$ m$^{-2}$ TeV$^{-1}$ at energy $E_0 = 1$~TeV, spectral index $\alpha = 2.111 \pm 0.057$, and cutoff energy $E_c=4$~TeV \citep{acciari11}.
Unlike previous observations, this TeV activity did not have
a counterpart at X-ray energies and was thus an ``orphan'' flare.

\begin{landscape}
\begin{figure}
\centering{ 
\includegraphics[scale=0.9]{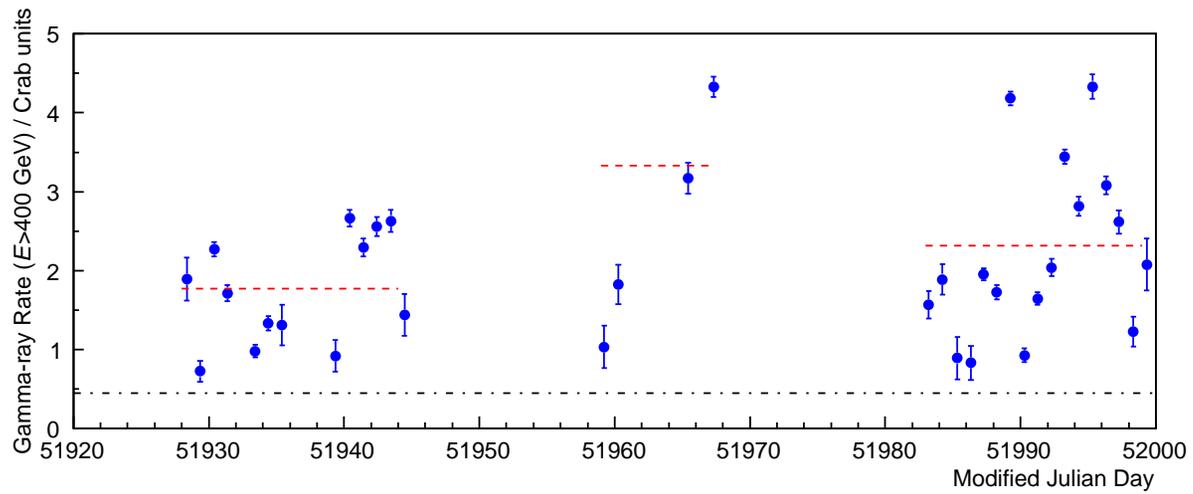}
\caption{Mrk421 gamma-ray rate ($E>$~400~GeV) for the period of high VHE activity from January 2001 to March 2001. The dashed lines represent the average rate (weighted mean) for each of the three darkruns in this period; the dot-dashed line indicates the mean rate for the full 14-year data set (1995--2009).}
\label{fig15}
}
\end{figure}
\end{landscape}

(iii) The most notable variability in the Mrk421 flux level
was observed in 2001, with an exceptionally
bright long-term gamma-ray high state lasting from January
to May (MJD 51928--52053). During the months of January and
March, the flux was highly variable and was four to five times 
the average flux; in
February, it increased even further (see Figure~\ref{fig15}). 
It is probably wrong to characterize this activity as a flare 
since it was more like a series of flares.
The source reached a maximum peak flux of $\sim$13 Crab
on 27th~February (MJD 51967). On one night during this
high TeV flux state, the
detection significance was $47\sigma$ for a four-hour exposure 
with an average flux during the four hours of $4.18\pm0.09$ Crab.

For the 2000--2001 observing season, the energy spectrum of Mrk421 
in the 0.38--8.2 TeV energy range
was measured on timescales of a month over a large range of flux
states \citep{krennrich02} and fitted using a power law with exponential cutoff
model \citep{krennrich02,aharonian05}. The position of the cutoff energy $E_c$ is highly correlated with
the photon index $\alpha$. To test for spectral variability, the cutoff 
energy was fixed at $E_c = 4.3$~TeV. Between November 2000 and April 2001, the TeV gamma-ray energy spectrum of Mrk421 hardened from a photon index 
$\alpha = 2.75 \pm 0.11$ at the onset of the large flare to
$\alpha = 1.89 \pm 0.04$ at the peak flux state in late February 2001
\citep{krennrich02} and then softened during the decay
of the flare event. This month-scale flare provides evidence
for a shift in the spectral energy distribution to higher energies
in the GeV-TeV band during the peak of the flaring episode. For
shorter flaring events, however, there is not always a clear correlation
between the TeV gamma-ray rate and the spectral shape. 

A detailed study was made of one week of data in March 2001 \citep{fossati08} 
during which there were several short TeV flares and there was 
significant simultaneous coverage in the X-ray and optical bands. 
During the flare on 19th~March (MJD 51987), the TeV signal was found to lag the 
X-ray signal in the 1--4~keV band by $35\pm12$~minutes. However,
the lag was less at higher X-ray energies. No correlation was found 
between the TeV and optical signals.

\section{Conclusions}

Atmospheric Cherenkov telescopes are ideally suited for studying 
AGN, but observations are most sensitive on clear dark nights. The large 
collection areas associated with the technique make them particularly
suited to the study of time variations. Because of their small 
fields of view, they are generally limited to the study 
of just one AGN at a time. The profusion of sources and the small 
number of sensitive telescopes available makes long-term monitoring of a 
particular source difficult. The Whipple 10~m telescope, while lacking the flux sensitivity of the major new observatories, H.E.S.S., MAGIC and VERITAS, was devoted over a 14-year timespan to observations of Mrk421 (along with a small number of other AGN) and thus provided a unique database. The Whipple 10~m gamma-ray telescope has since been mothballed in the summer of 2011, 
after 43 years of operation.

Mrk421 is the brightest and most studied blazar that is 
an emitter at TeV energies. The wealth of 
observational data makes it an ideal target for modeling the 
possible emission mechanisms. However, it is not 
yet clear to what extent it can be regarded as the archetypal 
TeV-emitting AGN. Given that a large number of blazars are 
only observed when they flare, it is useful 
to know the duty cycle of the flaring activity in order to 
estimate the chances of detection in any planned observing
campaign. Although emission at X-ray and TeV gamma-ray wavelengths
has been shown to be correlated, emission at other wavelengths (radio,
optical, infrared), which can be studied with ground-based
telescopes, is not strongly correlated. Hence 
it is difficult to plan a multiwavelength
campaign to observe flaring activity that could be triggered by
a non-gamma-ray signal.
 
Only a small number of blazars have been detected by the Whipple 
10~m telescope (Mrk421, Markarian 501 (Mrk501), 1ES2344+514, H1426+428 
and 1ES1959+650). Apart from Mrk421 and Mrk501, the average TeV flux 
level measured for these blazars is extremely low. For example, the 
average rate observed by the Whipple 10~m telescope from H1426+428 
during 2000 and 2001 was only ($10\pm2$)\% of the Crab Nebula rate 
\citep{horan02}. 1ES2344+514 and 1ES1959+650 were only detected by the 10~m telescope 
because of strong flaring activity \citep{catanese98,holder03}. 

It is therefore difficult to confirm the predictions of flaring 
frequency using 10~m data alone. As more TeV blazars are identified using data from H.E.S.S., 
MAGIC, and VERITAS, and their flaring statistics determined in long exposures, 
it will be possible to ascertain to what extent Mrk421 is typical
of TeV-emitting AGN.

\section*{Acknowledgements}

This research is supported by grants from the U.S. Department of 
Energy Office of Science, the U.S. National Science Foundation 
and the Smithsonian Institution, by NSERC in Canada, by Science 
Foundation Ireland (SFI 10/RFP/AST2748) and by STFC in the U.K. 
We acknowledge the excellent work of the technical support staff 
at the Fred Lawrence Whipple Observatory and at the collaborating 
institutions in the construction and operation of the instruments.

\clearpage

\clearpage

\clearpage

\clearpage

\clearpage

\clearpage

\clearpage

\clearpage

\clearpage

\clearpage

\clearpage

\clearpage

\clearpage

\clearpage

\clearpage

\clearpage

\end{document}